\documentclass[showpacs,amsmath,amssymb,twocolumn,prl,superscriptaddress]{revtex4-1}
\usepackage{amssymb}
\usepackage[dvips]{graphicx}
\usepackage{enumerate}
\usepackage{epsfig}
\usepackage{subfigure}
\usepackage{xcolor}
\usepackage[T1]{fontenc}
\usepackage{fullpage}
\usepackage{amsthm,amsfonts,amssymb,amscd,mathrsfs,xspace,framed}
\usepackage{amsmath}
\usepackage{color}
\usepackage{setspace}
\usepackage{url}
\usepackage{wrapfig}
\usepackage{enumitem}
\begin{document}

\newcommand{\Rv}[1]{{\sl{\textcolor{blue}{#1}}}}

\title{On-chip parallel processing of quantum frequency combs for high-dimensional hyper-entanglement generation}

\author{Liang Zhang}
\affiliation{James C. Wyant College of Optical Sciences, The University of Arizona, Tucson, Arizona 85721, USA}
\author{Chaohan Cui}
\affiliation{James C. Wyant College of Optical Sciences, The University of Arizona, Tucson, Arizona 85721, USA}
\author{Jianchang Yan}
\affiliation{Research and Development Center for Solid State Lighting, Institute of Semiconductors, Chinese Academy of Sciences, Beijing 100083, China}
\author{Yanan Guo}
\affiliation{Research and Development Center for Solid State Lighting, Institute of Semiconductors, Chinese Academy of Sciences, Beijing 100083, China}
\author{Junxi Wang}
\affiliation{Research and Development Center for Solid State Lighting, Institute of Semiconductors, Chinese Academy of Sciences, Beijing 100083, China}
\author{Linran Fan}
\email{lfan@optics.arizona.edu}
\affiliation{James C. Wyant College of Optical Sciences, The University of Arizona, Tucson, Arizona 85721, USA}

\maketitle

\textbf{
High-dimensional encoding and hyper-entanglement are unique features that distinguish optical photons from other quantum information carriers, leading to improved system efficiency and novel quantum functions. However, the disparate requirements to control different optical degrees of freedom have prevent the development of complete integrated platforms that is capable of leveraging the complementary benefits of high-dimensional encoding and hyper-entanglement at the same time.
Here we demonstrate the chip-scale solution to the generation and manipulation of high-dimensional hyper-entanglement. This is achieved by the parallel processing of multiple quantum frequency combs in the path domain. Cavity-enhanced parametric down-conversion with Sagnac configuration is implemented to ensure the spectral indistinguishability. Simultaneous entanglement in path and frequency is realized with high dimensions. On-chip reconfiguration of the entanglement structure is also demonstrated. We further present quantum interference in both entanglement degrees of freedom with high visibility. Our work provides the critical step for the efficient and parallel processing of quantum information with integrated photonics. 
} 

The generalization of two-level quantum systems to high dimensions provides the capability to verify quantum theories with stronger criteria, perform quantum computing with better error resilience, and conduct quantum communications with higher capacity and noise robustness~\cite{collins2002bell,cerf2002security,lanyon2009simplifying}. Optical photons are the ideal candidate for high-dimensional encoding~\cite{erhard2020advances}. Various degrees of freedom including orbital angular momentum~\cite{leach2010quantum,fickler2012quantum}, frequency~\cite{xie2015harnessing,reimer2016generation,jaramillo2017persistent,kues2017chip}, spatial~\cite{walborn2010spatial,dada2011experimental,edgar2012imaging}, and temporal modes~\cite{de2002creating,vagniluca2020efficient}, have been used for high-dimensional photonic entanglement generation. Besides high-dimensional encoding, simultaneous entanglement in different degrees of freedom (hyper-entanglement) is another promising route to enhance the quantum information processing capability with complementary features~\cite{deng2017quantum}. In particular, deterministic quantum logic~\cite{sheng2010deterministic} and complete Bell measurement~\cite{walborn2003hyperentanglement, schuck2006complete, barreiro2008beating} can be realized with the assistance of hyper-entanglement. By coherently combining high-dimensional encoding and hyper-entanglement, novel photonic technologies can be developed to realize critical quantum functions such as cluster state generation for universal quantum information processing~\cite{reimer2019high,imany2019high,ciampini2016path}.


Enhanced phase stability and high nonlinear efficiency make integrated photonics the ideal platform to implement complex large-scale quantum photonic systems~\cite{elshaari2020hybrid,wang2020integrated}. Among different degrees of freedom, path and frequency are two promising candidates for on-chip quantum information encoding. Path-encoding features the complete state manipulation with arbitrary unitary transformation at the expense of large-size photonic circuits. Frequency-encoding can generate large-scale quantum states with ultra-compact devices~\cite{reimer2016generation,jaramillo2017persistent,kues2017chip}. However, the difficulty in high-efficiency quantum frequency conversion prevents the demonstration of complex quantum state transformation on chip~\cite{kobayashi2016frequency,fan2016integrated, lu2018electro,joshi2020frequency}.

\begin{figure*}[htbp]
\centering
\includegraphics[width=16cm]{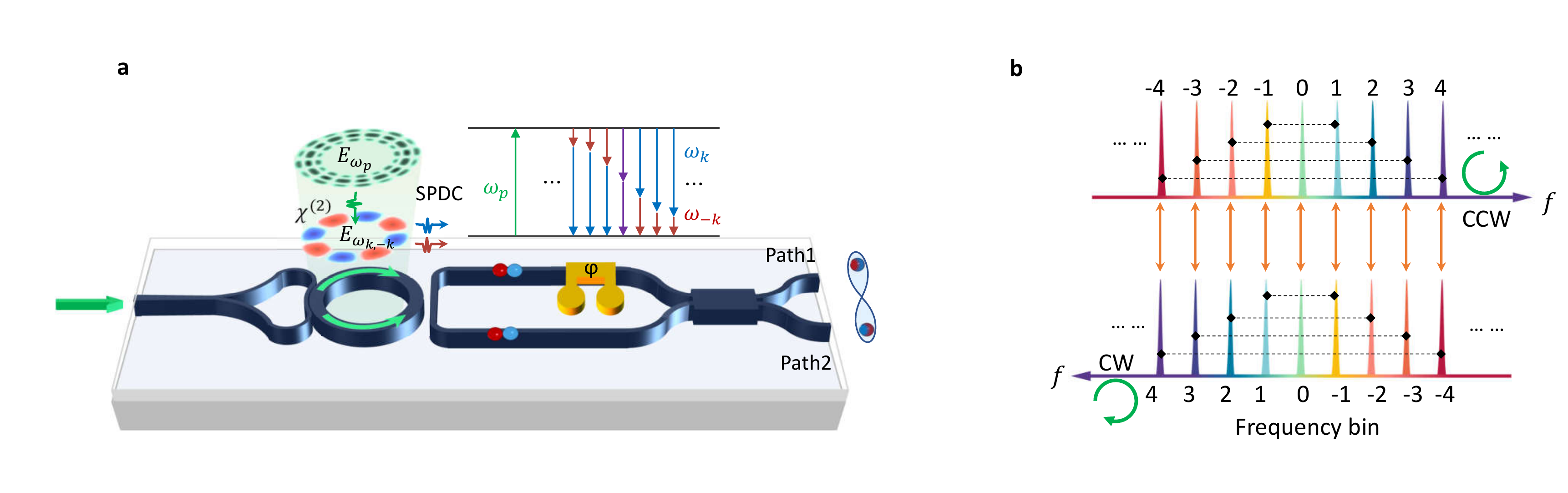}
\caption{\textbf{High-dimensional hyper-entanglement generation scheme}. \textbf{a}, Schematic to show the device layout for the generation of high-dimensional hyper-entanglement. The continuous-wave pump (green) is coupled onto the chip, equally splitted with a Y-junction, and coupled into the integrated ring cavity in both CW and CCW directions. Intra-cavity parametric down-conversion generates frequency-entangled photon pairs across multiple cavity resonances. Output photon pairs are interfered with tunable relative phase $\varphi$ on a 50/50 beamsplitter. \textbf{b}, Output entanglement structure with frequency correlation in symmetric spectral modes across CW and CCW paths.}
\label{fig:figure1}
\end{figure*}

Here, we propose and demonstrate the on-chip generation and control of high-dimensional hyper-entanglement leveraging the complementary features of path and frequency encoding. Multiple quantum frequency combs with near-unity spectral overlap are simultaneously generated with cavity-enhanced parametric down-conversion in Sagnac configuration. We realize the parallel processing of quantum frequency combs with uniform response and high visibility, leading to the precise reconfiguration of the hyper-entanglement structure. High-dimensional quantum interference is further performed to verify the high quality of the entanglement generation and manipulation.

\vspace{4pt}
\textbf{Hyper-entanglement generation scheme}

The schematic to generate high-dimensional hyper-entanglement is shown in Fig.~\ref{fig:figure1}a. The single-frequency pump in the visible regime is equally splitted, and coupled into the clockwise (CW) and counter-clockwise (CCW) directions of the integrated photonic ring cavity. Pockels nonlinearity enables efficient parametric down-conversion inside the cavity. With the large phase matching bandwidth, photon pairs in the telecom regime are generated in the superposition of multiple cavity resonances, forming quantum frequency combs. The simultaneous generation of multiple quantum frequency combs in the same cavity using Sagnac configuration ensures the near-identical spectral distribution. The relative phase $\varphi$ between frequency combs can be controlled by an on-chip phase shifter. After interference at the balanced multimode interferometer, the output state can be written as 
\begin{align}
\nonumber    |\psi_{out}\rangle=\frac{1}{\sqrt{2N+2}}\sum_{k=0}^{N}\sin\varphi (\hat{b}_k^{\dagger} \hat{b}_{-k}^{\dagger} -  \hat{c}_k^{\dagger} \hat{c}_{-k}^{\dagger}) &|0\rangle \\ 
    -\frac{1}{2\sqrt{2N+1}}\sum_{k=-N}^N\cos\varphi (\hat{b}_k^{\dagger} \hat{c}_{-k}^{\dagger} +  \hat{b}_{-k}^{\dagger} \hat{c}_{k}^{\dagger}) &|0\rangle
\label{Eq.1}
\end{align}
where $\hat{b}^{\dagger}$ and $\hat{c}^{\dagger}$ are the creation operators for the upper and lower paths, $k$ labels the frequency mode, and the total number of frequency modes is $2N+1$ (Supplementary Section I). The output state is entangled in the frequency domain due to the energy conservation. At the same time, the output state can be configured as a two-photon NOON state in the path domain if the relative phase $\varphi$ is set to zero~\cite{kok2002creation, pryde2003creation}. Therefore, the output state is entangled in both path and frequency, forming the high-dimensional hyper-entanglement  (Fig.~\ref{fig:figure1}b).

\begin{figure*}[htbp]
\centering
\includegraphics[width=16cm]{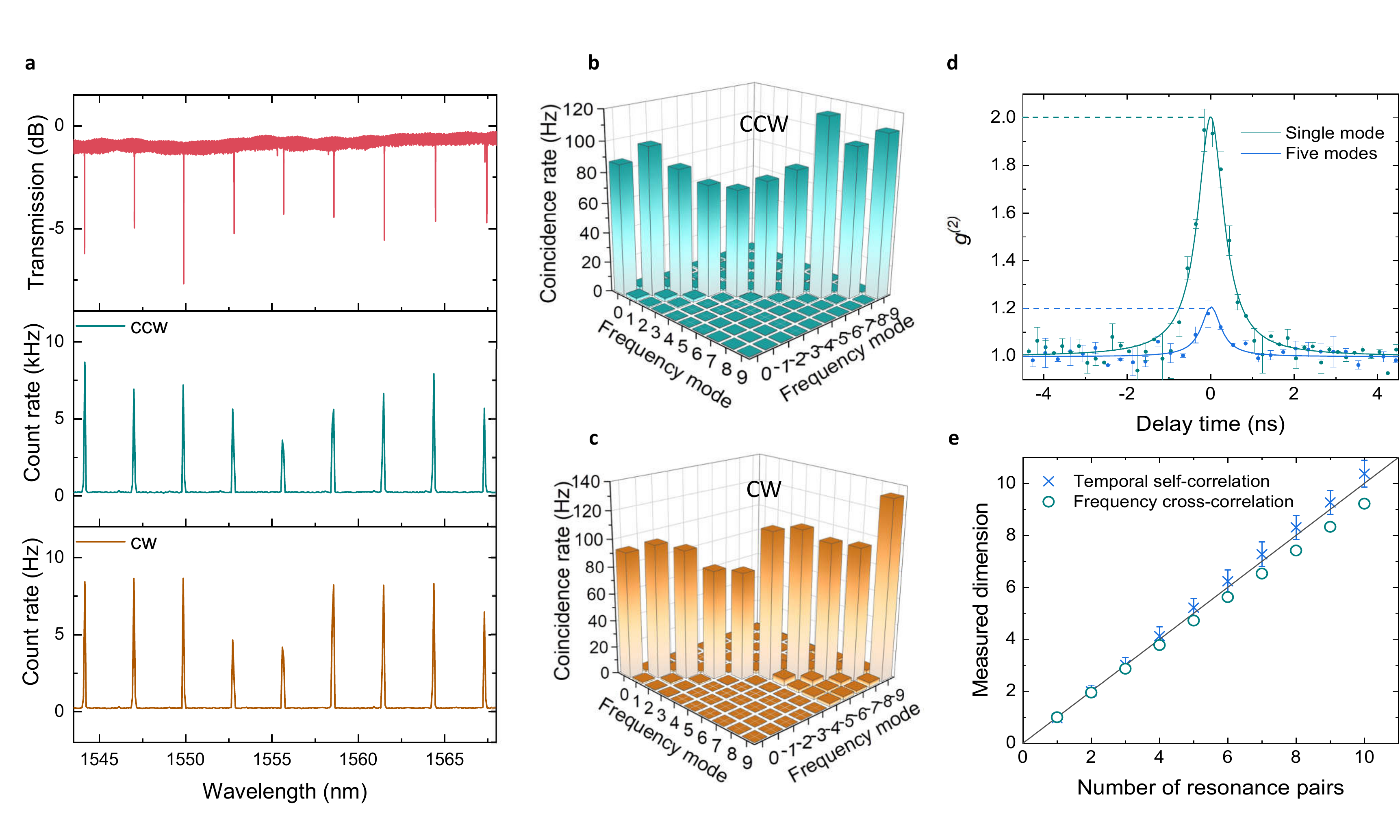}
\caption{ \textbf{Generation of quantum frequency combs}. \textbf{a}, Classical transmission spectrum and single-photon spectrum of CW and CCW quantum frequency combs. \textbf{b}, Measured frequency correlation of the CCW quantum frequency comb. \textbf{c}, Measured frequency correlation of the CCW quantum frequency comb. \textbf{d}, Self-correlation of photons generated from the single degenerate resonance (green) and five resonances (blue). \textbf{e}, Dimension of the CW quantum frequency comb with different numbers of resonance pairs. Upper (blue crosses) and lower (green circles) bounds are obtained from the temporal self-correlation measurement and the Schmidt decomposition of the frequency correlation matrix respectively.}
\label{fig:figure2}
\end{figure*}

\vspace{4pt}
\textbf{Pockels quantum frequency comb}

The complete photonic circuit is fabricated with aluminum nitride (AlN) on sapphire wafer. The photonic ring cavity has a radius of 60~$\mu$m with free-spectral range of 362~GHz. The phase matching for parametric down-conversion is realized with modal dispersion engineering (Supplementary Section II)~\cite{guo2017parametric}. We first performed experiments to characterize quantum frequency combs.  Single-photon spectrum is measured with narrow bandpass filters and superconducting nanowire single photon detectors (SNSPDs). Similar frequency mode locations and single-photon rate profiles are observed in both CW and CCW directions (Fig.~\ref{fig:figure2}a). Different frequency modes in the same quantum frequency comb are then separated by tunable filters to measure the correlation (Fig.~\ref{fig:figure2}b and c). Strong correlation is only observed between frequency modes located symmetrically to the degenerate wavelength at 1555.68~nm. The coincidence-to-accidental ratio above 25~dB has been achieved for all resonance pairs. The lower bound of the quantum state dimension can be obtained by the Schmidt decomposition of the frequency correlation matrix~\cite{law2000continuous, fedorov2014schmidt}. The Schmidt number scales linearly with the number of resonance pairs (Fig.~\ref{fig:figure2}e). With ten resonance pairs, the lower bound of the system dimension is estimated above 9.3 for each quantum frequency comb.

We further estimate the upper bound of the quantum state dimension through the self-correlation measurement in the time domain with Hanbury-Brown-Twiss setup (Supplementary Section III). The peak value of the self-correlation function at zero time delay $g^{(2)}(0)$ directly reflects the effective mode number $N_\mathrm{eff}$ $(g^{(2)}(0)=1+\frac{1}{N_\mathrm{eff}})$ (Fig.~\ref{fig:figure2}d)~\cite{christ2011probing}. The effective mode number for all resonances is approximately unity, showing the high purity of the generated quantum frequency comb (one effective mode per resonance)~\cite{helt2010spontaneous,cui2021high}. The upper bound of the total effective mode number can be estimated by the summation of individual separable modes (Fig.~\ref{fig:figure2}e). A linear relation between the upper bound of the system dimension and the resonance pairs is clearly observed. With ten resonance pairs, an upper bound of $10.39\pm0.52$ is achieved. As the lower and upper bounds are both rounded to ten, we show that the system dimension of each quantum frequency comb is ten.

\begin{figure*}[htbp]
\centering
\includegraphics[width=16cm]{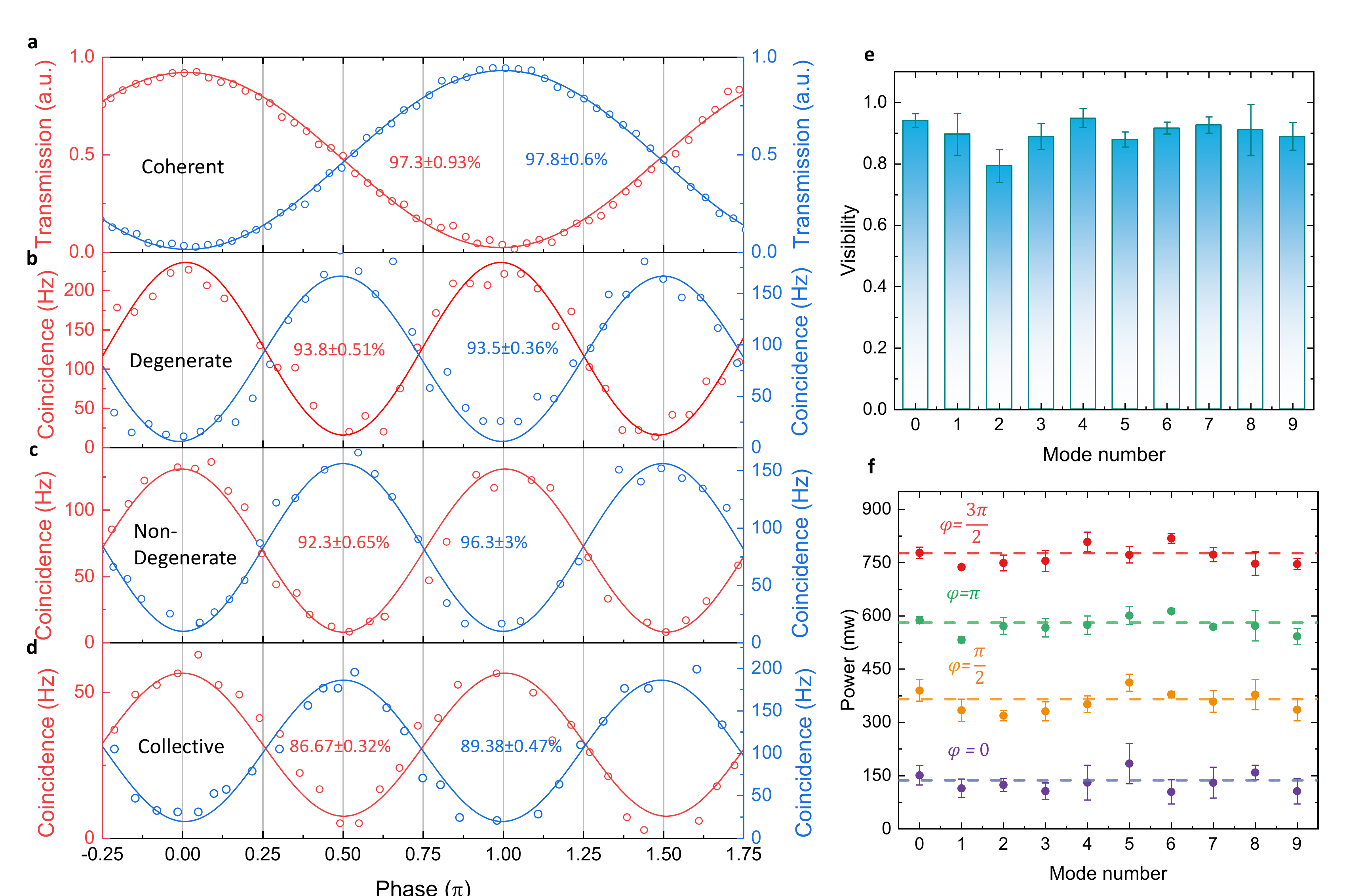}
\caption{\textbf{Path-domain entanglement and quantum interference}. \textbf{a}, Transmission of the coherent light through a Mach-Zehnder interferometer. \textbf{b}, Self-correlation (red) and cross-correlation (blue) of the degenerate photon pair with visibility $93.8\pm0.51\%$ and $93.5\pm0.36\%$ respectively. \textbf{c}, Correlation between photon pairs from non-degenerate resonance pairs (Mode 4) with the same path (red) and different paths (blue) with visibility $92.3\pm0.65\%$ and $96.3\pm3\%$ respectively. \textbf{d}, Correlation between photon pairs from five resonance pairs with the same path (red) and different paths (blue) with visibility $86.67\pm0.32\%$ and $89.38\pm0.47\%$. \textbf{e}, Visibility of quantum interference with different resonance pairs. \textbf{f}, Thermal-optic phase shifter condition for different phases and resonance pairs.}
\label{fig:figure3}
\end{figure*}

\vspace{4pt}
\textbf{Parallel processing in path domain}

We verify the path-domain performance through tunable on-chip quantum interference between CW and CCW quantum frequency combs. In contrast to classical interference where only phase $\varphi$ is obtained,  a $2\varphi$ phase shift is applied to the quantum frequency comb. The phase $\varphi$ is controlled with therm-optic effect through a local heater (Supplementary Section IV). The interference pattern of the coherent light through a Mach-Zehnder interferometer is used as the classical reference (Fig.~\ref{fig:figure3}a). We first isolate the degenerate resonance and measure the self-correlation of the upper path $S_{0,0}$ and the cross-correlation between the two paths $C_{0,0}$ (Fig.~\ref{fig:figure3}b). Both self- and cross-correlation show period half of the classical interference with complementary count rates. This agrees with Eq.~(\ref{Eq.1}) which suggests $S_{0,0}\propto \cos^2\varphi$ and $C_{0,0}\propto \sin^2\varphi$. The quantum interference shows high visibility of $93.8\pm0.51\%$ without background noise subtraction.

Another key difference between the classical and quantum interference is the dispersion response. The interference pattern shifts depending on wavelengths in the classical case. In the quantum case, the collective phase of the photon pair determines the interference pattern, which is the fundamental principle behind nonlocal dispersion cancellation~\cite{franson1992nonlocal}. With the fixed sum frequency, the total accumulated phase for symmetric resonances ($\varphi_k+\varphi_{-k}$) stays constant, leading to the same interference pattern for different resonance pairs. We verify this by measuring the cross-correlation between symmetric frequency modes in the same path $S_{k,-k}$, as well as in different paths $C_{k,-k}$ (Fig.~\ref{fig:figure3}c). The interference pattern stays unchanged, and high visibility is maintained  (Fig.~\ref{fig:figure3}e). In particular, the control condition of the phase shifter remains consistent for different resonance pairs (Fig. 3f). This enables the parallel processing of quantum frequency combs. The interference pattern and quality of quantum frequency combs are maintained the same as individual frequency modes (Fig.~\ref{fig:figure3}d).

\begin{figure*}[htbp]
\centering
\includegraphics[width=16cm]{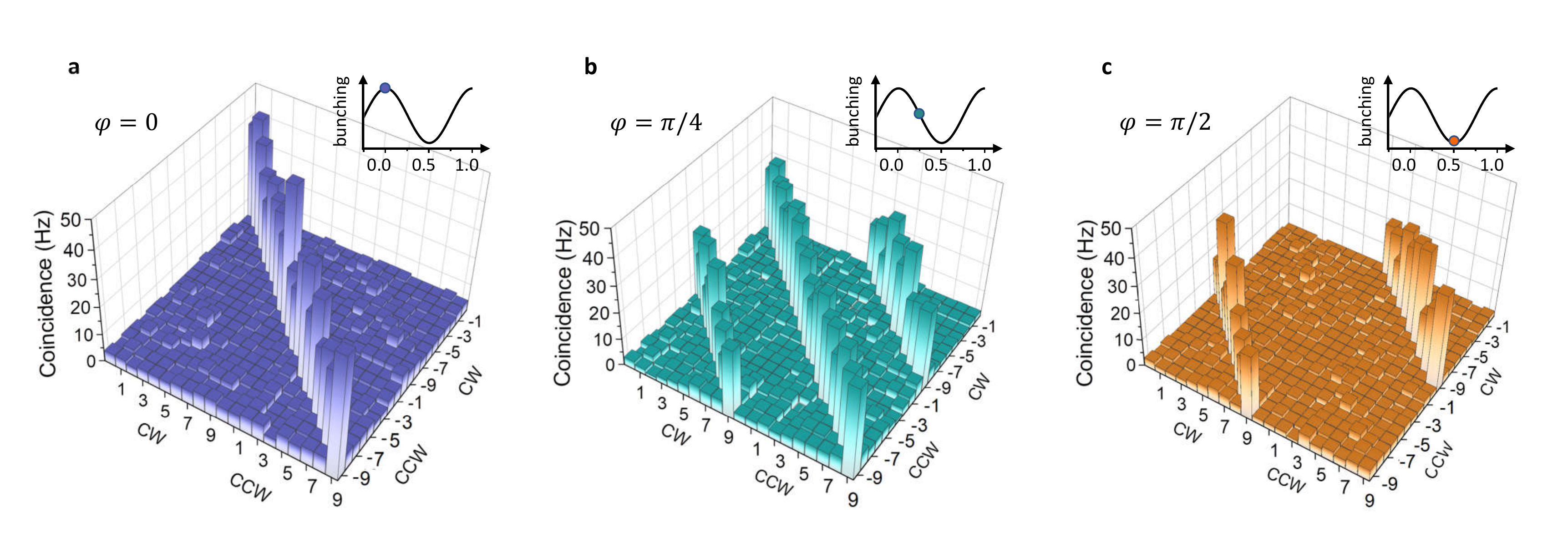}
\caption{\textbf{Quantum state reconfiguration.} Correlation matrix between different path and frequency modes under (\textbf{a}) bunching with $\varphi=0$, (\textbf{b}) superposition with $\varphi=\pi/4$, and (\textbf{c}) splitting with $\varphi=\pi/2$ conditions.}
\label{fig:figure4}
\end{figure*}

The parallel processing of quantum frequency combs can be further examined through the simultaneous measurement of the correlation in both path and frequency domains. Quantum frequency combs can be reconfigured to exhibit bunching, coherent superposition, and splitting between the two paths by setting the phase shifter to be $\varphi=0$, $\pi/4$, and $\pi/2$ respectively (Fig.~\ref{fig:figure4}a-c). Under the splitting condition, the CW and CCW quantum frequency combs undergo the reverse Hong-Ou-Mandel process, leading to the deterministic separation of the two photons~\cite{chen2007deterministic}. Therefore, no correlation is observed between any resonance pairs in the same path (Fig.~\ref{fig:figure4}a). Under the bunching condition, the two photons are grouped into the same path. Therefore, strong correlation can be observed within either path, but only between symmetric frequency modes. This indicates the generation of simultaneous entanglement in path and frequency domains. Under all photonic circuit settings, the coincidence-to-accidental ratio remains above 10 dB for all resonance pairs.

\vspace{4pt}
\textbf{High-dimensional quantum interference}

\begin{figure*}[htbp]
\centering
\includegraphics[width=16cm]{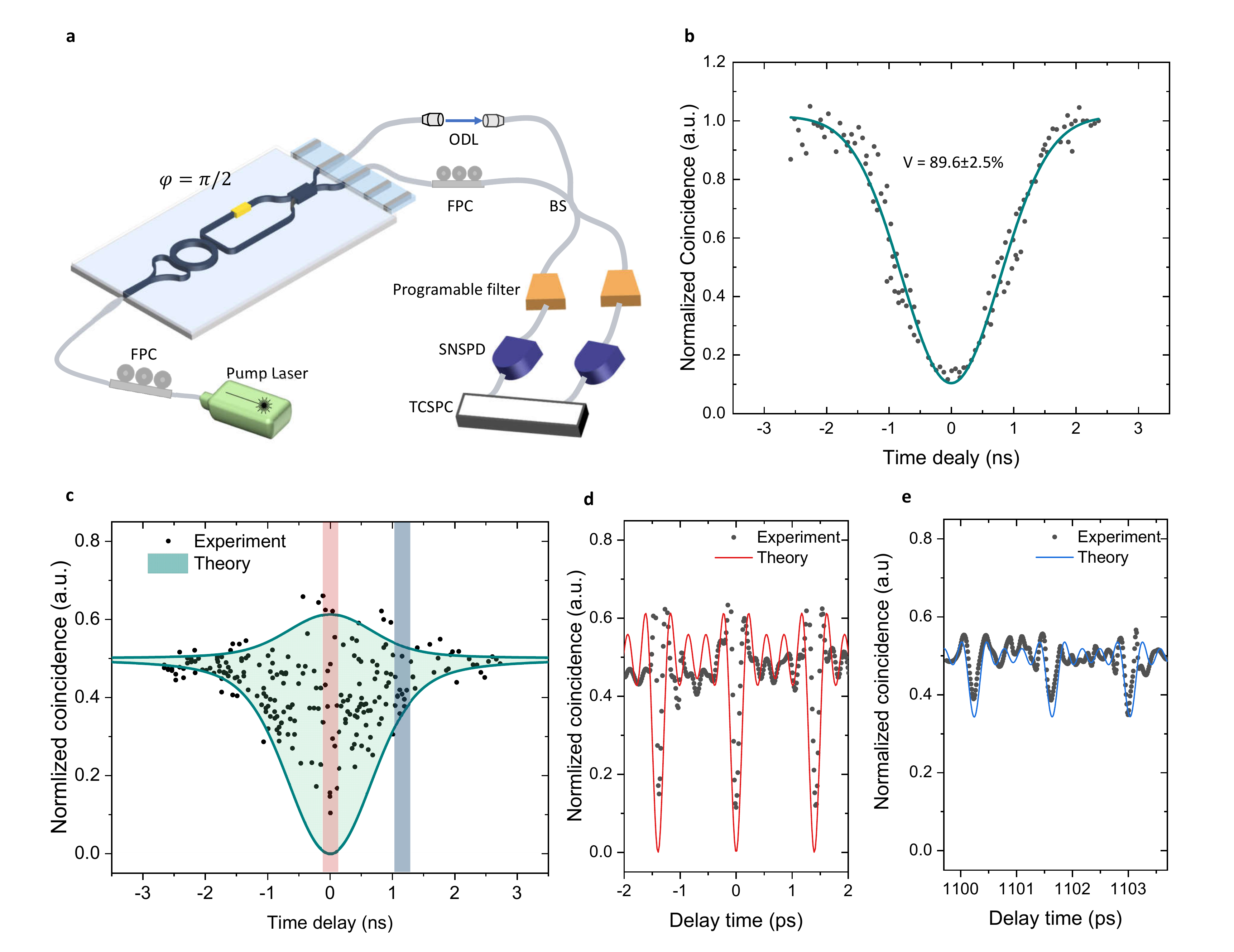}
\caption{ \textbf{High-dimensional quantum interference.} \textbf{a}, Experiment setup for quantum interference. FPC, fiber polarization controler; ODL, optical delay line; TCSPC, time-correlated single photon counter.
\textbf{b}, Normalized coincidence rate between the two SNSPDs with photon pairs in the degenerate resonance. \textbf{c}, Normalized coincidence rate between the two SNSPDs with photon pairs from five resonance pairs. Theoretical envelop of the coincidence rate is also plotted. \textbf{d}, Measured coincidence rate (black dot) and theoretical calculation (red) near zero time delay, corresponding to the red-shaded area in \textbf{c}. \textbf{e}, Measured coincidence rate (black dot) and theoretical calculation (blue) at time delay around 1.1 ns, corresponding to the blue-shaded area in \textbf{c}.}
\label{fig:figure5}
\end{figure*}

We further configure the output state under the splitting condition with $\varphi=\pi/2$ and perform the quantum interference between two quantum frequency combs with different time delays \cite{hong1987measurement} (Fig.~\ref{fig:figure5}a). A tunable optical delay line is placed in one path to control the relative time delay, and a fiber polarization controller is used to match the polarization between the two paths. A pair of programmable filters are used to shape the frequency profile in each path. After interference at a balanced fiber beamspliter, the coincidence is recorded by SNSPDs. We first select only the degenerate frequency mode in both paths. The standard Hong-Ou-Mandel interference is observed with a single dip at zero time delay (Fig.~\ref{fig:figure5}b). Visibility around $89.6\pm2.5\%$ is achieved without subtracting the background noise. The dip width is estimated around $1.57\pm0.04$~ns, which matches the linewidth of the degenerate frequency mode (Supplementary Section II). We further program the filters to incorporate the first five resonance pairs. We observe the fast oscillation of the coincidence under different time delays due to the beating between different resonance pairs (Fig.~\ref{fig:figure5}c). Upon zoom-in of the coincidence pattern at zero time delay, narrow coincidence dips with high visibility can be observed, proving the intra-pair phase coherence between symmetric resonances (Fig.~\ref{fig:figure5}d)~\cite{lingaraju2019quantum}. The revival of the coincidence dip is also observed with delay period of 1.4~ps, which agrees with the free-spectral range of the photonic ring cavity (362 GHz). The visibility of the coincidence dips decreases with non-zero time delay due to the finite temporal coherence of quantum frequency combs (Fig.~\ref{fig:figure5}e). The coherence length is determined by the cavity loss. Therefore, the envelop of coincidence dips matches the standard Hong-Ou-Mandel interference with a single frequency mode (Fig.~\ref{fig:figure5}c). 

\vspace{4pt}
\textbf{Discussion}

In addition to parametric down-conversion for efficient quantum state generation, Pockels nonlinearity can also provide strong electro-optic effect. High-frequency optical modulation and efficient pulse shaping have been demonstrated using integrated AlN photonic devices with performance exceeding commercial fiber-based and free-space modulators~\cite{fan2019spectrotemporal}. The coherent combination of different frequency modes can be realized with efficient phase modulation, which can verify the inter-pair phase coherence of quantum frequency combs~\cite{kues2017chip,lingaraju2019quantum}. With on-chip programmable filters, the complete coherent transformation of the high-dimensional hyper-entanglement state in both path and frequency domains can be realized on chip based on integrated Pockels nonlinearity~\cite{khan2010ultrabroad,lukens2017frequency}.

Scalability and functionality are the ultimate criteria to evaluate any integrated quantum photonic systems. We have provided a new prospective by proposing and demonstrating the coherent combination of path- and frequency-encoding on the same chip. Parametric down-conversion with nanophotonic cavity enhancement leads to the efficient generation of quantum frequency combs. High-dimensional hyper-entanglement has been generated by the parallel processing of quantum frequency combs. We further present the on-chip high-fidelity state control and high-visibility quantum interference. The unique Pockels nonlinearity can enable the integration of complete path-frequency control on a single integrated platform. This work can significantly expand the dimension of quantum photonic systems while simultaneously enhancing the quantum functionality, providing new opportunities for photonic quantum information processing.

\vspace{4pt}
\textbf{Acknowledgments}
LZ, CC, and LF acknowledge the support from U.S. Department of Energy, Office of Advanced Scientific Computing Research, (Field Work Proposal ERKJ355); Office of Naval Research (N00014-19-1-2190); National Science Foundation (ECCS-1842559). LF acknowledges the support from II-VI foundation. Device is fabricated in the OSC cleanroom at the University of Arizona, and the cleanroom of Arizona State University. SNSPDs are supported by NSF MRI INQUIRE.

\vspace{4pt}
\textbf{Author contributions}

The experiments were conceived by LF and CC. The device was designed and fabricated by LZ and LF. Measurements were performed by LZ, LF, and CC. Analysis of the results was
conducted by LZ, LF, and CC. The AlN wafer was provided by JY, YG, and JW. All authors participated in the manuscript preparation.

\vspace{4pt}
\textbf{Methods}

\textbf{AlN wafer growth.}
The AlN wafer is grown on sapphire substrate with c-plane (0001) by MOCVD. Trimethylaluminum (TMA), ammonia (NH$_3$) and hydrogen (H$_2$) are used as the aluminum source, nitrogen source, and carrier gas, respectively. The growth started with a 25-nm low-temperature AlN buffer layer at 550~$^o$C, followed by the growth of 1-$\mu$m high quality AlN layer at 1200~$^o$C.

\textbf{Device fabrication.}
The device is fabricated with 1-$\mu$m AlN film on sapphire grown by MOCVD. We use electron-beam lithography to exposure FOX-16 resist to define the photonic circuit. After development in Tetramethylammonium hydroxide (TMAH), the pattern is transferred to AlN layer with reactive ion etching (RIE) using Cl$_2$/BCl$_3$/Ar chemistry. Then 2-$\mu$m SiO$_2$ is deposited by PECVD as the cladding layer. Finally, electrodes for phase shifter is defined with S1813 photoresist using photolithography, followed by Ti/Pt (5/200nm) deposition and lift-off in Acetone.

\vspace{4pt}
\textbf{Data availability}

The data that support the findings of this study are available from the corresponding author upon reasonable request

\bibliography{Ref}
\end{document}